\RequirePackage{fixltx2e}
\documentclass[aps,prb,reprint,floatfix]{revtex4-1}
\usepackage{amsmath,amsfonts,amssymb} %
\usepackage{mathtools} %
\usepackage{wasysym} %
\usepackage{bm} %
\usepackage[bbgreekl]{mathbbol} %
\usepackage{graphicx} %
\usepackage[dvipsnames,table]{xcolor} %
\usepackage{tabularx} %
\usepackage[acronym]{glossaries} %
\usepackage{braket} %
\usepackage{multirow} % Required for multicells

\usepackage{fancyhdr} %
\usepackage{microtype} %
\usepackage{natmove}

\usepackage[byname]{smartref} %
\usepackage[breaklinks, %
colorlinks, linkcolor=Blue, citecolor=Blue, urlcolor=Blue]{hyperref} %
\usepackage[version=3]{mhchem}
\AtBeginDocument{\usepackage{booktabs}} % Fix incompatibility with

\newcommand{\PP}{\mathcal{P}} %
\newcommand{\RR}{\mathbf{R}} %
\newcommand{\BC}{\color{black} } %
\newcommand{\eq}[1]{Eq.~(\ref{#1})} %
\newcommand{\fig}[1]{Fig.~\ref{#1}} %

\def\be{\begin{equation}} %
\def\ee{\end{equation}} %
\def\bea{\begin{eqnarray}} %
\def\eea{\end{eqnarray}} %

\newacronym{QPE}{QPE}{quantum phase estimation} %
\newacronym{VQE}{VQE}{variational quantum eigensolver} %
\newacronym{UCC}{UCC}{unitary coupled cluster} %
\newacronym{QCC}{QCC}{qubit coupled cluster} %
\newacronym{FCI}{FCI}{full configurational interaction} %
\newacronym{CASCI}{CASCI}{complete active space configurational
  interaction} %
\newacronym{JW}{JW}{Jordan--Wigner} %
\newacronym{BK}{BK}{Bravyi--Kitaev} %
\newacronym[longplural={degrees of freedom}, %
firstplural={degrees of freedom (DOF)}, plural={DOF}]{DOF}{DOF}{degree
  of freedom} %
\newacronym[longplural={equations of motion}, %
firstplural={equations of motion (EOM)}, %
plural={EOM}]{EOM}{EOM}{equation of motion} %
\newacronym{PES}{PES}{potential energy surface} %
\newacronym{CI}{CI}{configuration interaction} %
\newacronym{QMF}{QMF}{qubit mean-field} %
\newacronym{SQP}{SQP}{sequential quadratic programming} %
\newacronym{RHF}{RHF}{restricted Hartree--Fock}

\begin{document}
  
\author{Artur F. Izmaylov${}^{a,b}$} 
\affiliation{${}^a$Chemical Physics Theory Group, Department of Chemistry,
  University of Toronto, Toronto, Ontario, M5S 3H6, Canada; 
  ${}^b$ Department of Physical and Environmental Sciences,
  University of Toronto Scarborough, Toronto, Ontario, M1C 1A4,
  Canada; ${}^c$Department of Quantum Field Theory, Taras Shevchenko National University of Kyiv, Kyiv, 03022, Ukraine}
    
\author{Tzu-Ching Yen${}^a$} 
\affiliation{${}^a$Chemical Physics Theory Group, Department of Chemistry,
  University of Toronto, Toronto, Ontario, M5S 3H6, Canada; 
  ${}^b$ Department of Physical and Environmental Sciences,
  University of Toronto Scarborough, Toronto, Ontario, M1C 1A4,
  Canada; ${}^c$Department of Quantum Field Theory, Taras Shevchenko National University of Kyiv, Kyiv, 03022, Ukraine}

  \author{Robert A. Lang${}^{a,b}$} 
\affiliation{${}^a$Chemical Physics Theory Group, Department of Chemistry,
  University of Toronto, Toronto, Ontario, M5S 3H6, Canada; 
  ${}^b$ Department of Physical and Environmental Sciences,
  University of Toronto Scarborough, Toronto, Ontario, M1C 1A4,
  Canada; ${}^c$Department of Quantum Field Theory, Taras Shevchenko National University of Kyiv, Kyiv, 03022, Ukraine}

\author{Vladyslav Verteletskyi${}^{a,b,c}$} 
\affiliation{${}^a$Chemical Physics Theory Group, Department of Chemistry,
  University of Toronto, Toronto, Ontario, M5S 3H6, Canada; 
  ${}^b$ Department of Physical and Environmental Sciences,
  University of Toronto Scarborough, Toronto, Ontario, M1C 1A4,
  Canada; ${}^c$Department of Quantum Field Theory, Taras Shevchenko National University of Kyiv, Kyiv, 03022, Ukraine}
  
 \title{Unitary partitioning approach to the measurement problem in the Variational Quantum Eigensolver method}

\date{\today}

\begin{abstract}
To obtain estimates of electronic energies, the Variational Quantum Eigensolver (VQE) 
technique performs separate measurements for multiple parts of the system Hamiltonian. 
Current quantum hardware is restricted to projective single-qubit measurements, and thus, only parts 
of the Hamiltonian which form mutually qubit-wise commuting groups can be measured simultaneously. 
The number of such groups in the electronic structure Hamiltonians grows as $N^4$, where  
$N$ is the number of qubits, and thus puts serious restrictions on the size of the systems that can be studied. 
Using a partitioning of the system Hamiltonian as a linear combination of unitary operators we found 
a circuit formulation of the VQE algorithm that allows one to measure a group of fully 
anti-commuting terms of the Hamiltonian in a single series of single-qubit measurements.  
Numerical comparison of the unitary partitioning to previously used grouping of Hamiltonian 
terms based on their qubit-wise commutativity shows an $N$-fold reduction in the number of measurable groups.   
\end{abstract}

\glsresetall

\maketitle

%%%%%%%%%%%%%
% Reviewers: N. Rubin, A. Mezzacampo, P. Love, J. Whitfield,  

\section{Introduction}

The \gls{VQE} method\cite{Peruzzo:2014/ncomm/4213, Yung:2014/SR/3589,Jarrod:2016/njp/023023, Wecker:2015/pra/042303,Olson:2017ud,McArdle:2018we} provides a practical approach 
to solving the eigen-value problem for many-body interacting  
Hamiltonians on current and near-future universal quantum computers.  
\gls{VQE} is a hybrid quantum-classical approach based on the variational theorem and a mapping of the 
electronic structure problem 
\bea
\hat H_e (\RR) \ket{\Psi(\RR)} = E_e(\RR) \ket{\Psi(\RR)}
\eea
to its qubit counterpart 
\bea
\hat H_q (\RR)\ket{\Psi_q(\RR)} = E_e(\RR) \ket{\Psi_q(\RR)}.
\eea
Here, $\hat H_e (\RR)$ is the electronic Hamiltonian, $\RR$ is the nuclear configuration of interest, 
$E_e(\RR)$ is the electronic energy,
$\hat H_q(\RR)$ is the qubit Hamiltonian obtained from a second quantized form of $\hat H_e (\RR)$\cite{Helgaker:2000} 
using one of the fermion-qubit mappings,\cite{Bravyi:2002/aph/210, Seeley:2012/jcp/224109,Tranter:2015/ijqc/1431, Setia:2017/ArXiv/1712.00446,Havlicek:2017/pra/032332} and $\ket{\Psi_q(\RR)}$ is the corresponding 
qubit wave-function. For notational simplicity, in what follows, we will skip the nuclear configuration but will always 
assume its existence as a parameter. 

In VQE, the quantum computer prepares a trial qubit wavefunction $\ket{\Psi_q}$ 
and then does measurements to accumulate statistics for the expectation value of the qubit Hamiltonian. 
The classical computer completes the VQE cycle by 
suggesting a new trial wavefunction based on previous expectation values of energy. The two steps, on 
classical and quantum computers, are iterated until convergence. 
One of the strengths of the VQE approach is ability to use relatively short-depth quantum circuits to 
construct a qubit wavefunction $\ket{\Psi_q}$ close to the true eigenstate of the problem.
Note though that the VQE scheme cannot measure the whole system Hamiltonian at once, because the system 
Hamiltonian is not the Hamiltonian of qubits and is not physically implemented in the quantum computer. 
This is one of the differences between universal quantum computing and quantum simulation.\cite{cirac:2012,cirac:2018} 

Measuring parts of the system Hamiltonian is a very time-consuming task. 
Experimentally, one can only measure single-qubit Pauli operators, 
$\hat \sigma_i = \hat x_i,\hat y_i$ or  $\hat z_i$. A regular qubit Hamiltonian       
\begin{equation}
  \label{eq:Hq}
  \hat H_q = \sum_I C_I\,\hat P_I
\end{equation}
is a linear combination of products of Pauli operators $\hat P_I$ (Pauli ``words'') 
for different qubits,  
\begin{equation}
  \label{eq:Pi}
  \hat P_I = \prod_{i=1}^{N} \hat \sigma_{i}^{(I)},
\end{equation}
where $\hat \sigma_i^{(I)}$ is one of the $\hat x,\hat y,\hat z$ Pauli operators or the identity $\hat 1$ operator
for the $i^{\rm th}$ qubit, and $N$ is the total number of qubits.
For single-qubit measurements one can group only those terms that share a common tensor product 
eigen-basis. Thus, during the measurement, 
the system wavefunction can collapse to a set of unentangled eigenstates common to 
all Pauli operators in the group.  A simple criterion for grouping terms based on shared tensor product 
eigen-basis is their mutual commutativity within single-qubit subspaces or qubit-wise commutativity.\cite{VVpap1}   

 If the canonical molecular orbitals are used as an orthonormal basis to setup 
the fermionic second-quantized Hamiltonian, the total number of terms in the qubit Hamiltonian scales as the fourth power of the number of qubits needed to represent the electronic wavefunction. A better scaling alternative for 
an orthonormal fermionic basis are plane-waves,\cite{Babbush:PRX2018} where the number of the
qubit Hamiltonian terms only scales quadratically with the number of plane waves. The downside of this 
approach is a large prefactor for the quadratic scaling that makes the crossing point between 
the quadratic and quartic scalings far out of reach of current and near future quantum computers.

Recently, we have proposed an efficient technique for grouping qubit-wise commuting (QWC) terms of the Hamiltonian
by mapping the qubit Hamiltonian to a graph where connectivity between vertices representing the Hamiltonian 
terms indicates qubit-wise commutativity.\cite{VVpap1} The grouping problem then can be 
reformulated as a problem of graph partitioning to the minimum number of fully connected 
subgraphs. This problem is known as the minimum clique cover problem,\cite{GraphBook:2004} 
and it is NP-hard.\cite{Karp:1972} Using polynomial 
heuristics, it was found that this grouping approach can reduce the total number of 
simultaneously measurable parts by a factor of 3 (on average) from the total number of terms in \eq{eq:Hq}, which still leaves a large number of groups to be measured.\cite{VVpap1}  

Another way to reduce the number of separately measured groups has been suggested recently in Ref.~\citenum{Izmaylov:2019gb}, where the idea of the single-qubit measurement was generalized to the case 
when the result of one qubit measurement was used to determine what single-qubit operator needs to be 
measured next. Partitioning of the qubit Hamiltonian into fragments that can be measured with such 
feed-forward measurement procedures increased the number of terms that can be grouped together and thus 
reduced the number of separately measured groups. However, 
even though such feed-forward measurements were demonstrated in some 
experiments,\cite{nArriagada:2018ju,Prevedel:2007ca,Moqanaki:2015iw,Reimer:2018cv} they have not 
yet became available in mainstream quantum computing hardware available to the public.  
Another difficulty with this approach is that a procedure for ensuring the optimality of this partitioning has yet to be found.

Here, we explore a different route to the Hamiltonian partitioning, which is
based on the idea that if the Hamiltonian were a unitary operator, its expectation value could be obtained 
in one set of single-qubit measurements. Although the qubit Hamiltonian is not a single unitary operator, its individual 
Pauli products in \eq{eq:Hq} are. It is possible to combine some Pauli products  
to larger groups of unitary operators that constitute measurable sets. 
The grouping condition for Pauli words involves anti-commutativity of terms rather than their commutativity 
and thus present an alternative approach to the measurement problem.
Optimal grouping of unitary fragments is possible through solving a minimum clique cover problem for an  
anti-commutativity graph constructed based on the qubit Hamiltonian.

The rest of the paper is organized as follows. In Sec.~\ref{sec:UP} we develop a partitioning of the
qubit Hamiltonian to a minimal number of unitary fragments. Section~\ref{sec:UOMC}
details the quantum computing circuit for measuring expectation values for these 
unitary fragments. Assessment of the new scheme is done on a set of molecular systems 
with the number of terms in $\hat H_q$ up to fifty thousands (Sec.~\ref{sec:results}).   
Section~\ref{sec:conclusions} summarizes the main results and provides concluding remarks.

\section{Theory}
\label{sec:theory}

\subsection{Unitary Partitioning} 
\label{sec:UP}

Here we will discuss how to partition the qubit Hamiltonian into a linear combination of 
the minimum number of unitary operators   
\bea\label{eq:HU}
\hat H_q = \sum_{n=1}^{M} d_n \hat U_n,
\eea
where $\{d_n\}_{n=1}^M$ is a set of real coefficients, and $\hat U_n$ are $M$ unitary operators.

Note that all Pauli words are hermitian unitary operators, $\hat P_I^{\dagger} \hat P_I = \hat P_I^2 = 1$.
However, a general sum of unitary operators is non-unitary
\bea
\left(\sum_I C_I\hat P_I\right)^{\dagger} \left(\sum_I C_I\hat P_I\right) \ne 1. 
\eea
To make $\sum_I C_I\hat P_I$ unitary, it is sufficient to impose the following three additional conditions: 
1) $Im(C_I^*C_J)=0$, 2) $\sum_I |C_I|^2=1$, and 3) $\{\hat P_I,\hat P_J\} = 2\delta_{IJ}$ (where $\{.,.\}$ 
is the anti-commutator).
The first two conditions are easy to satisfy for any partial sum of the Hamiltonian in \eq{eq:Hq} because all 
coefficients are real, hence, only their renormalization is required
\bea
\sum_I C_I\hat P_I = C \sum_I \frac{C_I}{C}\hat P_I,~ C= \left( \sum_I C_I^2\right)^{1/2},
\eea 
then the first two conditions for unitarity will be satisfied for the sum with coefficients $C_I/C$. 

To satisfy the third condition, one needs to partition the Hamiltonian into groups of Pauli matrices 
that mutually anti-commute. To reduce the number of unitary operators needed to represent 
$\hat H_q$ in \eq{eq:HU}, we would like to maximize the number of mutually anti-commuting terms
in each group. Recently, it was found that a similar problem of 
 finding minimum partitioning into groups of mutually QWC terms can be solved 
 using a graph representation for the Hamiltonian.\cite{VVpap1} There, every Pauli word was considered as a
 graph vertex and edges were put between the terms that qubit-wise commute. The grouping problem is 
 equivalent to the very well-known minimum clique cover (MCC) problem.\cite{GraphBook:2004} 
 For the anti-commuting relation,
 one can also build a graph representation of the Hamiltonian where two Pauli word vertices are 
 connected if the corresponding operators anti-commute. Since two Pauli words always 
 either commute or anti-commute, the anti-commutativity graph is complementary to the 
 commutativity graph. The MCC provides the minimum number of fully connected 
 subgraphs, cliques. Each clique forms a set of anti-commuting Pauli words whose linear combination 
 can represent a unitary operator 
 \bea
\hat U_n &=& \frac{1}{d_n} \sum_I C_I^{(n)}\hat P_I^{(n)}, \\
d_n &=& \left( \sum_I (C_I^{(n)})^2\right)^{1/2},
 \eea 
where $\{\hat P_I^{(n)},\hat P_J^{(n)}\}=2\delta_{IJ}$.  
Thus, for the further discussion we will assume that solving the MCC problem for the Hamiltonian anti-commutativity graph 
 provides the minimum number of $\hat U_n$ operators. 
 One caveat in approaching the optimal partitioning problem by solving the MCC problem is 
that the latter is NP-hard.\cite{Karp:1972} To avoid the exponential scaling in solving MCC, we resort to various 
polynomial heuristic algorithms, which were discussed in Ref.~\citenum{VVpap1}.

\subsection{Unitary Operator Measuring Circuit} 
\label{sec:UOMC}
 
Partitioning of the $H_q$ in Eq.~(\ref{eq:HU}) allows us to rewrite the energy expectation value as
\bea
\bar{E} = \bra{\Psi} \hat H_q \ket{\Psi} = \sum_n d_n \bra{\Psi} \hat U_n \ket{\Psi}. 
\eea
Accounting for a unitary preparation of the wavefunction $\ket{\Psi} = \hat U\ket{\bar{0}}$, 
where $\ket{\bar{0}}$ is $N$ qubit vacuum or initial all-qubits-up state. For measuring,
it is convenient to rewrite $\bar{E}$ in a symmetric form as 
\bea
\bar{E} = \frac{1}{2}\sum_n d_n (\bra{\Psi} \hat U_n \ket{\Psi} + \bra{\Psi} \hat U_n^{\dagger} \ket{\Psi}).
\eea
By introducing $\ket{\Phi_n} =\hat U^\dagger \hat U_n \hat U\ket{\bar{0}}$ states the energy estimate 
can be written as 
\bea
\bar{E} = \frac{1}{2}\sum_n d_n (\braket{\bar{0}\vert \Phi_n} + \braket{\Phi_n\vert\bar{0}}).
\eea
In what follows we will discuss how to measure the individual components 
\bea 
\braket{\bar{0}\vert \Phi_n} + \braket{\Phi_n\vert\bar{0}} = 2 Re \braket{\bar{0}\vert \Phi_n},
\eea 
which are directly connected to the energy estimate:
\bea
\bar{E} = \sum_n d_n Re\braket{\bar{0}\vert \Phi_n}. 
\eea

\begin{figure}[h!]
  \centering %
  \includegraphics[width=1\columnwidth]{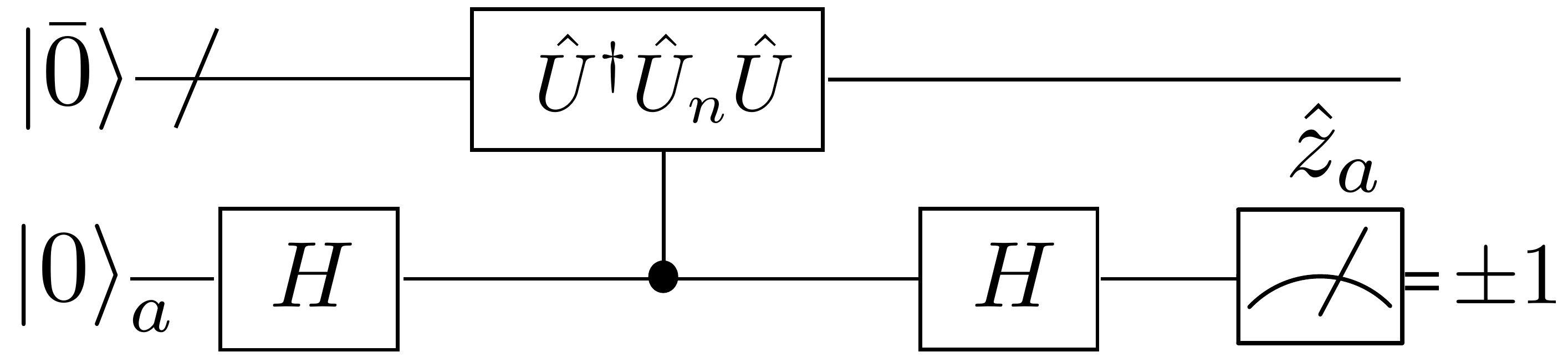}
  \caption{ Circuit for extracting values of $Re\braket{\bar{0}\vert \Phi_n}$, it requires $N+1$ qubits and 
  $M$ series of the $\hat z_a$ single-qubit measurements. }
  \label{fig:sch}
\end{figure}

To measure the real part of the overlap $\braket{\bar{0}\vert \Phi_n}$ we will not use the swap 
test\cite{PhysRevLett.87.167902} 
because this test produces the absolute value of the overlap instead of its real part. Our
approach to evaluating $Re\braket{\bar{0}\vert \Phi_n}$ will be as follows (see \fig{fig:sch}). 
The initial state is a tensor product $\ket{\bar{0}}\otimes\ket{0}_a$ of one ancilla and $N$ target  
qubits. 
First, the Hadamard gate $H = (\hat x+\hat z)/\sqrt{2}$ is applied to the ancilla qubit
\bea
\ket{\Psi_1} = \ket{\bar{0}}\otimes(\ket{0}_a+\ket{1}_a)/\sqrt{2}. 
\eea 
Second, using a controlled unitary operator $\hat U^\dagger\hat U_n\hat U$ 
the following superposition is created
\bea
\ket{\Psi_2} = (\ket{\bar{0}}\otimes\ket{0}_a+\ket{\Phi_n}\otimes\ket{1}_a)/\sqrt{2}.
\eea
Third, another Hadamard gate rotates the $\ket{\Psi_2}$ state into
\bea
\ket{\Psi_3} = \frac{1}{2}[\ket{\Phi_{n+}}\otimes\ket{0}_a+\ket{\Phi_{n-}}\otimes\ket{1}_a],
\eea
where $\ket{\Phi_{n\pm}} = \ket{\bar{0}}\pm\ket{\Phi_n}$. 
 Finally, the expectation value of the $\hat z_a$ operator is measured. After 
obtaining the statistics for these measurements one can extract $Re\braket{\bar{0}\vert \Phi_n}$ 
from the probabilities of the $z_a =\pm 1$ measurement outcomes, $p_{\pm 1} = (1\pm Re\braket{\bar{0}\vert \Phi_n})/2$. 

\subsection{Circuit depth analysis}

How can one implement the controlled $\hat U^\dagger\hat U_n\hat U$ transformation on a quantum computer? 
Any $\hat U$ can be presented as a product of one- and two-qubit operators for a regular VQE 
circuit to generate a trial wavefunction.\cite{Ryabinkin:2018/qcc}  
If $\hat U_n = \sum_{k=1}^{L}c_k \hat P_k$, 
where $ \sum_{k=1}^{L}c_k^2=1$, using the anti-commutativity of terms, this sum 
can be presented as a product of $2L-1$ exponents of Pauli words (entanglers) 
\bea
\label{eq:U_product}
\hat U_n = \prod_{k=1}^{L} e^{i\theta_k\hat P_k /2} \prod_{k=L}^{1}  e^{i\theta_k\hat P_k/2},
\eea
where $\theta_k$'s can be connected with $c_k$'s as 
\begin{align}
\theta_k = \arcsin \frac{c_k}{\sqrt{\sum_{j=1}^k  c_j^2}}.
\end{align} 
This connection is easy to understand from a geometric point of view for 
$c_k$'s as Cartesian coordinates of a point on a unit $L-1$-dimensional sphere and 
$\theta_k$'s as corresponding hyper-spherical coordinate components.  
Therefore, compared to $\hat U$,
the new transformation $\hat U^\dagger\hat U_n\hat U$
 in the worst case (no significant cancellation between terms in the product) will have 
twice as many terms in addition to $2L-1$ terms generated from $\hat U_n$. 
The $2L-1$ entanglers are not necessarily one- and two-qubit operators, 
 but one can implement each of the entanglers using $O(N)$ CNOT gates and single qubit rotations. \cite{Seeley:2012/jcp/224109} Consequently, the cost of $\hat U_n$ scales as $O(NL)$.

To implement the controlled $\hat U^\dagger\hat U_n\hat U$, all one-qubit {\BC operators} can be replaced by 
controlled-$U$ gates. For the two-qubit operators, we can find decompositions in 
CNOT and one-qubit gates\cite{PhysRevA.69.010301}, 
which are then replaced by Toffoli and controlled-$U$ gates. 
Hence, implementing the controlled $\hat U^\dagger\hat U_n\hat U$ is not asymptotically more 
expensive than implementing $\hat U^\dagger\hat U_n\hat U$.

 \subsection{Application to the projection formalism}
 
To impose physical symmetries one can construct projectors on irreducible representations of the symmetry 
group or algebra. These projectors can be always presented as a linear combination of unitary 
operators\cite{Yen:2019prj}
\bea\label{eq:Pexp}
\hat \PP = \sum_{k} a_k \hat U_k,
\eea 
and can be applied in the expectation values of the projected Hamiltonian 
\bea
\bar{E} &=& \frac{\bra{\Psi}\hat \PP^\dagger \hat H_q \hat \PP\ket{\Psi}}{\bra{\Psi}\hat \PP^\dagger \hat \PP\ket{\Psi}} \\
&=& \frac{\bra{\Psi} \hat H_q \hat \PP\ket{\Psi}}{\bra{\Psi}\hat \PP\ket{\Psi}}.
\eea
Here, the last equation used hermiticity, idempotency, and commutativity with the Hamiltonian for the 
symmetry projector. The expansions in unitary transformations 
for the projector [\eq{eq:Pexp}] and the Hamiltonian [\eq{eq:HU}] can be easily combined because 
a product of two unitary operators is unitary. Even though introducing the projector expansion will 
increase the number of terms for the measurement, it allows one to reduce the complexity of the unitary 
transformation for the preparation of $\ket{\Psi}$ by satisfying symmetry requirements by construction.\cite{Yen:2019prj} 

\begin{table*}[!htbp]
  \caption{The number of qubits ($N$), Pauli words in qubit Hamiltonians (Total), QWC groups ($M_{\rm QWC}$), 
  and unitary groups ($M$) produced by different heuristics for systems 
   with up to 14 qubits. The STO-3G basis has been used for all Hamiltonians unless specified otherwise.}
  \label{tab:small}
  \centering
    \begin{ruledtabular}
   \begin{tabular}{@{}lcccccccccccc@{}}
    Systems & $N$ & Total & $M_{\rm QWC}$  & GC & LF & SL & DS & RLF & DB & C & R & BKT\\
    \hline
    H$_2$ (BK)& 4 & 15 & 3 & 11 & 11 & 11 & 11 & 11 & 11 & 11 & 11 & 11 \\
    LiH  (Parity)& 4 & 100 & 25 & 33 & 33 & 23 & 29 & 19 & 18 & 20 & 21 & 16\\    
    H$_2$O (6-31G, BK)& 6 & 165 & 34 & 41 & 43 & 41 & 43 & 32 & 31 & 34 & 35 & 31 \\
    BeH$_2$  (BK) & 14 & 666 & 172 & 141 & 130 & 118 & 120 & 112 & 109 & 116 & 123 & -\\
BeH$_2$  (JW)& 14 & 666 & 203 & 139 & 135 & 121 & 119 & 110 & 108 & 120 & 128 & -\\
H$_2$O  (BK) & 14 & 1086 & 308 & 176 & 197 & 147 & 154 & 127 & 129 & 145 & 155 & -\\
H$_2$O (JW) & 14 & 1086 & 322 & 181 & 197 & 159 & 153 & 127 & 128 & 153 & 154 & -\\
     \end{tabular}
       \end{ruledtabular}
\end{table*}

\begin{table*}[!htbp]
   \caption{Comparison of RLF results for BK and JW transformed Hamiltonians: the number of unitary groups 
   ($M$), their maximum size (Max Size), and standard deviation of their size distributions (STD). The total number of Hamiltonian terms (Total) is almost everywhere the same for JW and BK; for the last two systems, JW numbers are in parenthesis.}
  \label{tab:varst}
    \begin{ruledtabular}
   \begin{tabular}{@{}lcccccccc@{}}
 \multirow{2}{*}{Systems} & \multirow{2}{*}{$N$} & \multirow{2}{*}{Total} & \multicolumn{3}{c}{BK} & \multicolumn{3}{c}{JW}\\
 \cline{4-6}\cline{7-9}
	 & & & $M$ & Max Size & STD & $M$ &Max Size& STD \\
    \hline
    BeH$_2$ / STO-3G & 14 & 666 & 112 & 10 & 2.0 & 110 & 11 & 2.1\\
	H$_2$O / STO-3G & 14 & 1086 & 127 & 13 & 2.0 & 127 & 15 & 2.2\\
	NH$_3$ / STO-3G & 16 & 3609 & 251 & 25 & 3.9 & 251 & 25 & 3.8\\
	N$_2$ / STO-3G & 20 & 2951 & 266 & 17 & 2.5 & 268 & 19 & 2.6\\
    BeH$_2$ / 6-31G & 26 & 9204 & 556 & 26 & 4.5 & 558 & 29 & 4.6\\
H$_2$O / 6-31G & 26 & 12732 & 767 & 33 & 5.2 & 779 & 32 & 5.2 \\
    NH$_3$ / 6-31G & 30 & 52758 (52806) & 1761 & 50 & 7.8 & 1781 & 50 & 7.8 \\
    N$_2$ / 6-31G & 36 & 34639 (34655) & 1402 & 43 & 5.9 & 1399 & 46 & 5.9 \\
     \end{tabular}
       \end{ruledtabular}
\end{table*}
		
\section{Numerical studies and discussion}
\label{sec:results}

To assess our developments we apply them to several molecular electronic Hamiltonians 
(Tables~\ref{tab:small} and \ref{tab:varst}). 
Details of generating these Hamiltonians are given in Supplementary Information. Some 
of these systems were used to illustrate performance of quantum computing techniques
previously.\cite{Kandala:2017/nature/242,Hempel:2018/prx/031022,Ryabinkin:2018/qcc}.

To solve the MCC problem we have used several heuristic algorithms based on either 
reformulating the problem as graph coloring or approximating it as finding and removing maximum cliques.\cite{VVpap1}  
The description of used heuristics can be found in Ref.\citenum{VVpap1} and original papers: 
Greedy Coloring (GC),\cite{Rigetti_doc} Largest First (LF),\cite{Welsh:1967} 
Smallest Last (SL),\cite{Matula:1972} DSATUR (DS),\cite{Brelaz:1973} Recursive 
Largest First (RLF),\cite{Leighton:1979} Dutton and Brigham (DB),\cite{dutton_brigham_1981} 
COSINE,\cite{hertz_1990}  Ramsey (R),\cite{boppana} Bron-Kerbosch-Tomita (BKT).\cite{tomita_tanaka_takahashi_2006}
All these heuristics except BKT have polynomial computational scaling  
with respect to the number of graph vertices.   

Table~\ref{tab:small} summarizes results of the unitary partitioning and compares it with previously 
used qubit-wise commutativity partitioning. 
Even though the BKT approach shows superior performance for the first three systems in 
Table~\ref{tab:small}, due to its exponential computational scaling, it cannot be used for larger systems.
Among polynomial algorithms, RLF is the best heuristic in terms of both computational time 
and the number of produced cliques, the latter is 20\% lower than that of the next-best algorithm. 
Thus, the RLF algorithm can be recommended for larger systems and has been applied for them 
(see Table~\ref{tab:varst}). Both maximum clique size and standard deviation of clique sizes 
grows approximately linearly with the number of qubits. The difference between results for 
JW and BK Hamiltonians are negligible.

\begin{figure}[h!]
  \centering %
  \includegraphics[width=1\columnwidth]{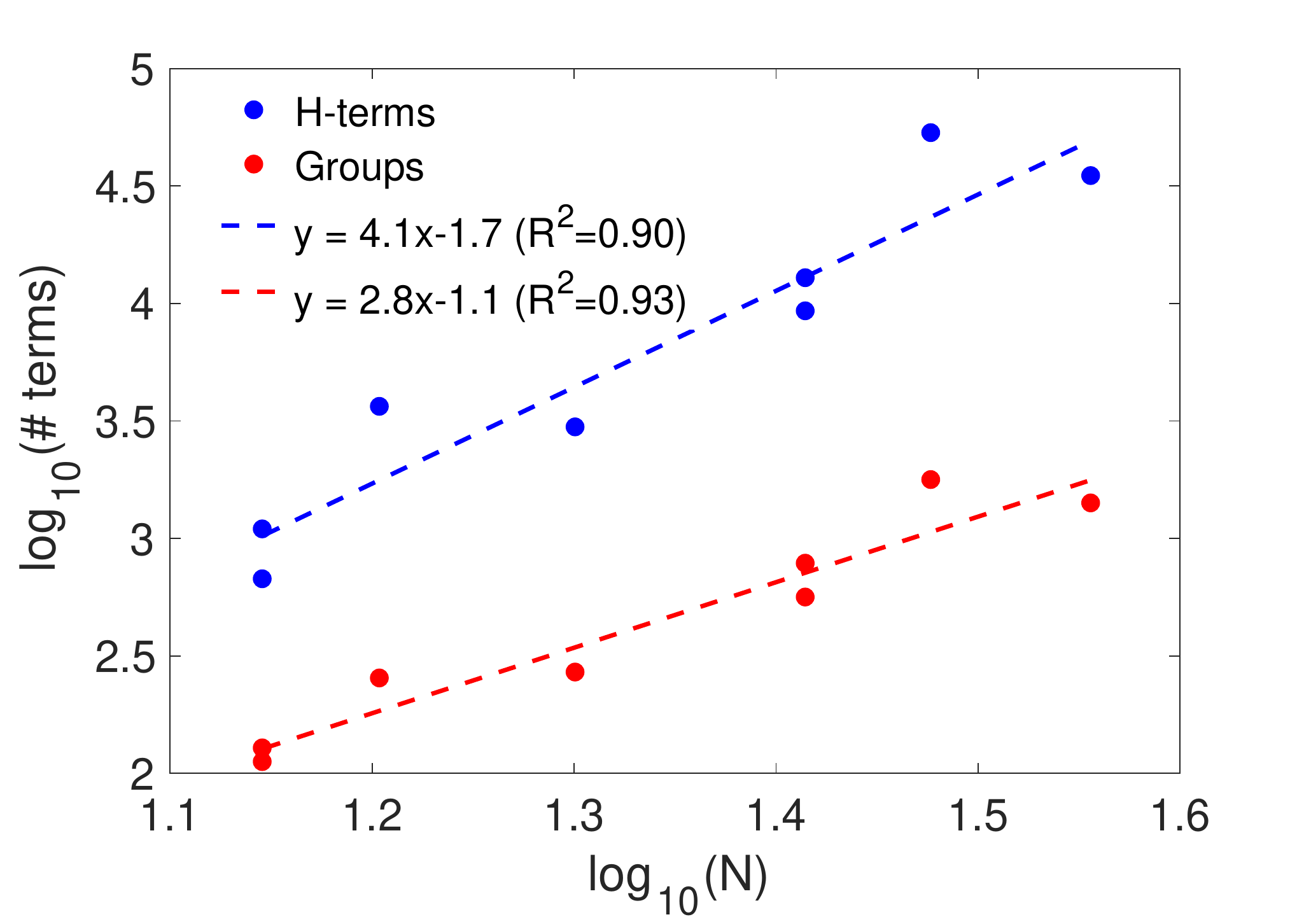}
  \caption{ Dependencies of the total number of the Hamiltonian terms (blue) 
  and the number of unitary groups (red) on the number of qubits for the systems in 
  Table~\ref{tab:varst} in the double log-scale. For the last two entries of Table~\ref{tab:varst}, 
  the JW and BK results were averaged.}
  \label{fig:scal}
\end{figure}

Fitting the number of the Hamiltonian terms and the number of unitary fragments on 
the number of qubits ($N$) for 
the systems of Table~\ref{tab:varst} in the double log-scale reveals $N^4$ and $N^3$ scalings respectively 
(\fig{fig:scal}).  
Thus, introducing unitary grouping reduces the number of terms to measure by a factor of $N$. 
Analytical proof of this result has been obtained recently in Ref.~\citenum{Zhao:2019vz}.
This reduction can be rationalized 
from the graph connectivity point of view. It is easy to show that an average
Pauli word has exponentially many more connections for the graph based on anti-commutativity 
compared to that based on QWC.

\section{Conclusions}
\label{sec:conclusions}

We have introduced and studied a new method for partitioning of the qubit Hamiltonian
to a linear combination of unitary transformations. This unitary partitioning allows us to reduce the number of 
separate measurements required in the VQE approach to the electronic structure problem. 
Testing the new technique on a set of molecular electronic Hamiltonians revealed  
an $N$-fold reduction in the number of operators that require separate measurements.
The unitary partitioning scheme increases depth of quantum circuits and introduces an 
additional ancilla qubit. For measuring 
a group of anti-commuting terms containing $L$ elements on a trial wavefunction prepared 
using $K$ entanglers, the depth of a new circuit becomes at least $2K+2L-1$ entanglers. 

The partitioning of the qubit Hamiltonian is done by representing it as a graph where every 
vertex corresponds to a single Pauli word and the edges are connecting the terms that are 
anti-commuting. In this representation, the problem of grouping terms that can form a unitary 
operator corresponds to finding fully connected subgraphs (cliques). To obtain optimal partitioning 
the number of groups should be the fewest. This is a well-known problem in discrete math, 
the minimum clique cover problem, which is solved using polynomial heuristic algorithms. 

Among various tested heuristics, the RLF approach is found to be the most efficient polynomial 
algorithm producing the lowest number of unitary groups. Hamiltonians
produced using different fermion-qubit transformations (JW and BK) had similar 
compression rates due to the unitary partitioning.    
 
Another advantage of the unitary partitioning is its straightforward incorporation of the symmetry 
projections that can always be presented as linear combinations of unitary operators. 

\section*{Acknowledgement}
A.F.I. is grateful to Prof. Michael Molloy for useful discussion and 
acknowledges financial support from Zapata Computing Inc., the Natural Sciences and
Engineering Research Council of Canada (NSERC), and the Mitacs Globalink Program. 
R.A.L acknowledges financial support from NSERC.

\section*{Note Added:} After submission of this manuscript to arXiv we became aware of 
several new proposals addressing the measurement problem, which appeared within a 
week or two from each other.\cite{MoscaA,ThomsonA,BabbushA,ChicagoA} 

\section*{Supplementary Information: Hamiltonian Generation}

\paragraph*{H$_2$ molecule:}
One- and two-electron integrals in the canonical \gls{RHF} molecular
orbitals basis for $R$(H-H)=1.5 \AA, 
were used in the \gls{BK} transformation to produce the corresponding qubit Hamiltonian.
Spin-orbitals were alternating in the order $\alpha$, $\beta$,
$\alpha$, .... 

\paragraph*{LiH molecule:}
Using the parity transformation for the LiH molecule at $R{\rm (Li-H)}=3.2$ \AA, 
a 6-qubit Hamiltonian containing 118 Pauli words was generated.
Spin-orbitals were arranged as ``first all $\alpha$ then all $\beta$'' in
the fermionic form; since there are 3 active molecular orbitals in the
problem, this leads to 6-qubit Hamiltonian. This qubit Hamiltonian has 3$^{rd}$ and 6$^{th}$
stationary qubits, which allowed us to replace the corresponding
$\hat z$ operators by their eigenvalues, $\pm 1$, thus defining the
different ``sectors'' of the original Hamiltonian. Each of these
sectors is characterized by its own 4-qubit effective Hamiltonian. The
ground state lies in the $z_{3} = -1$, $z_{6} = 1$ sector; 
the corresponding 4-qubit effective Hamiltonian ($\hat H_{\rm LiH}$) has
100 Pauli words.  

\paragraph*{\ce{H2O} molecule:}
6- and 26-qubit Hamiltonians were generated for this system in the 6-31G basis, and the
14-qubit Hamiltonian was generated using the STO-3G basis. 
The geometry for all Hamiltonians was chosen to be $R{\rm (O-H)}=0.75$ \AA~ and $\angle HOH = 107.6^{\circ}$
The 14- and 26-qubit Hamiltonians were obtained in OpenFermion using both JW and BK transformations without any modifications, while for the 6-qubit Hamiltonian we used several qubit reduction techniques detailed below. 

Complete active space (4,\,4) electronic Hamiltonian was converted to the qubit form using the \gls{BK}
  transformation grouping spin-orbitals as ``first all alpha than all beta''. The resulting 8-qubit Hamiltonian
  contained 185 Pauli terms. $4^\text{th}$ and $8^\text{th}$ qubits were found to be stationary;
  the ground state solution is located in the $z_{3} = 1$, $z_{7} = 1$ subspace. By integrating out 
  $z_{3}$ and $z_{7}$, the 6-qubit reduced Hamiltonian with 165 terms was derived.  

\paragraph*{\ce{N2}, \ce{BeH2}, and \ce{NH3} molecules:} 
The BK and JW transformations of the electronic Hamiltonian in the 6-31G and STO-3G
bases produced qubit Hamiltonians by OpenFermion. The nuclear geometry was fixed at  
$R{\rm (N-N)}=1.1$ \AA (\ce{N2}); $R{\rm (Be-H)} = 1.4 $ \AA, collinear geometry (\ce{BeH2});  
$\angle HNH = 107^{\circ}$ and  $R{\rm (N-H)} = 1.0 $ \AA (\ce{NH3}).

%\bibliography{qcomp-snap,books,programs,ekt-snap,databases,gp}

\begin{thebibliography}{43}%
\makeatletter
\providecommand \@ifxundefined [1]{%
 \@ifx{#1\undefined}
}%
\providecommand \@ifnum [1]{%
 \ifnum #1\expandafter \@firstoftwo
 \else \expandafter \@secondoftwo
 \fi
}%
\providecommand \@ifx [1]{%
 \ifx #1\expandafter \@firstoftwo
 \else \expandafter \@secondoftwo
 \fi
}%
\providecommand \natexlab [1]{#1}%
\providecommand \enquote  [1]{``#1''}%
\providecommand \bibnamefont  [1]{#1}%
\providecommand \bibfnamefont [1]{#1}%
\providecommand \citenamefont [1]{#1}%
\providecommand \href@noop [0]{\@secondoftwo}%
\providecommand \href [0]{\begingroup \@sanitize@url \@href}%
\providecommand \@href[1]{\@@startlink{#1}\@@href}%
\providecommand \@@href[1]{\endgroup#1\@@endlink}%
\providecommand \@sanitize@url [0]{\catcode `\\12\catcode `\$12\catcode
  `\&12\catcode `\#12\catcode `\^12\catcode `\_12\catcode `\%12\relax}%
\providecommand \@@startlink[1]{}%
\providecommand \@@endlink[0]{}%
\providecommand \url  [0]{\begingroup\@sanitize@url \@url }%
\providecommand \@url [1]{\endgroup\@href {#1}{\urlprefix }}%
\providecommand \urlprefix  [0]{URL }%
\providecommand \Eprint [0]{\href }%
\providecommand \doibase [0]{http://dx.doi.org/}%
\providecommand \selectlanguage [0]{\@gobble}%
\providecommand \bibinfo  [0]{\@secondoftwo}%
\providecommand \bibfield  [0]{\@secondoftwo}%
\providecommand \translation [1]{[#1]}%
\providecommand \BibitemOpen [0]{}%
\providecommand \bibitemStop [0]{}%
\providecommand \bibitemNoStop [0]{.\EOS\space}%
\providecommand \EOS [0]{\spacefactor3000\relax}%
\providecommand \BibitemShut  [1]{\csname bibitem#1\endcsname}%
\let\auto@bib@innerbib\@empty
%</preamble>
\bibitem [{\citenamefont {Peruzzo}\ \emph {et~al.}(2014)\citenamefont
  {Peruzzo}, \citenamefont {McClean}, \citenamefont {Shadbolt}, \citenamefont
  {Yung}, \citenamefont {Zhou}, \citenamefont {Love}, \citenamefont
  {Aspuru-Guzik},\ and\ \citenamefont {O'Brien}}]{Peruzzo:2014/ncomm/4213}%
  \BibitemOpen
  \bibfield  {author} {\bibinfo {author} {\bibfnamefont {A.}~\bibnamefont
  {Peruzzo}}, \bibinfo {author} {\bibfnamefont {J.}~\bibnamefont {McClean}},
  \bibinfo {author} {\bibfnamefont {P.}~\bibnamefont {Shadbolt}}, \bibinfo
  {author} {\bibfnamefont {M.-H.}\ \bibnamefont {Yung}}, \bibinfo {author}
  {\bibfnamefont {X.-Q.}\ \bibnamefont {Zhou}}, \bibinfo {author}
  {\bibfnamefont {P.~J.}\ \bibnamefont {Love}}, \bibinfo {author}
  {\bibfnamefont {A.}~\bibnamefont {Aspuru-Guzik}}, \ and\ \bibinfo {author}
  {\bibfnamefont {J.~L.}\ \bibnamefont {O'Brien}},\ }\href@noop {} {\bibfield
  {journal} {\bibinfo  {journal} {Nat. Commun.}\ }\textbf {\bibinfo {volume}
  {5}},\ \bibinfo {pages} {4213} (\bibinfo {year} {2014})}\BibitemShut
  {NoStop}%
\bibitem [{\citenamefont {Yung}\ \emph {et~al.}(2014)\citenamefont {Yung},
  \citenamefont {Casanova}, \citenamefont {Mezzacapo}, \citenamefont {McClean},
  \citenamefont {Lamata}, \citenamefont {Aspuru-Guzik},\ and\ \citenamefont
  {Solano}}]{Yung:2014/SR/3589}%
  \BibitemOpen
  \bibfield  {author} {\bibinfo {author} {\bibfnamefont {M.-H.}\ \bibnamefont
  {Yung}}, \bibinfo {author} {\bibfnamefont {J.}~\bibnamefont {Casanova}},
  \bibinfo {author} {\bibfnamefont {A.}~\bibnamefont {Mezzacapo}}, \bibinfo
  {author} {\bibfnamefont {J.}~\bibnamefont {McClean}}, \bibinfo {author}
  {\bibfnamefont {L.}~\bibnamefont {Lamata}}, \bibinfo {author} {\bibfnamefont
  {A.}~\bibnamefont {Aspuru-Guzik}}, \ and\ \bibinfo {author} {\bibfnamefont
  {E.}~\bibnamefont {Solano}},\ }\href@noop {} {\bibfield  {journal} {\bibinfo
  {journal} {Sci. Rep.}\ }\textbf {\bibinfo {volume} {4}},\ \bibinfo {pages}
  {3589} (\bibinfo {year} {2014})}\BibitemShut {NoStop}%
\bibitem [{\citenamefont {McClean}\ \emph {et~al.}(2016)\citenamefont
  {McClean}, \citenamefont {Romero}, \citenamefont {Babbush},\ and\
  \citenamefont {Aspuru-Guzik}}]{Jarrod:2016/njp/023023}%
  \BibitemOpen
  \bibfield  {author} {\bibinfo {author} {\bibfnamefont {J.~R.}\ \bibnamefont
  {McClean}}, \bibinfo {author} {\bibfnamefont {J.}~\bibnamefont {Romero}},
  \bibinfo {author} {\bibfnamefont {R.}~\bibnamefont {Babbush}}, \ and\
  \bibinfo {author} {\bibfnamefont {A.}~\bibnamefont {Aspuru-Guzik}},\ }\href
  {\doibase 10.1088/1367-2630/18/2/023023} {\bibfield  {journal} {\bibinfo
  {journal} {N. J. Phys.}\ }\textbf {\bibinfo {volume} {18}},\ \bibinfo {pages}
  {023023} (\bibinfo {year} {2016})}\BibitemShut {NoStop}%
\bibitem [{\citenamefont {Wecker}\ \emph {et~al.}(2015)\citenamefont {Wecker},
  \citenamefont {Hastings},\ and\ \citenamefont
  {Troyer}}]{Wecker:2015/pra/042303}%
  \BibitemOpen
  \bibfield  {author} {\bibinfo {author} {\bibfnamefont {D.}~\bibnamefont
  {Wecker}}, \bibinfo {author} {\bibfnamefont {M.~B.}\ \bibnamefont
  {Hastings}}, \ and\ \bibinfo {author} {\bibfnamefont {M.}~\bibnamefont
  {Troyer}},\ }\href {\doibase 10.1103/PhysRevA.92.042303} {\bibfield
  {journal} {\bibinfo  {journal} {Phys. Rev. A}\ }\textbf {\bibinfo {volume}
  {92}},\ \bibinfo {pages} {042303} (\bibinfo {year} {2015})}\BibitemShut
  {NoStop}%
\bibitem [{\citenamefont {Olson}\ \emph {et~al.}(2017)\citenamefont {Olson},
  \citenamefont {Cao}, \citenamefont {Romero}, \citenamefont {Johnson},
  \citenamefont {Dallaire-Demers}, \citenamefont {Sawaya}, \citenamefont
  {Narang}, \citenamefont {Kivlichan}, \citenamefont {Wasielewski},\ and\
  \citenamefont {Aspuru-Guzik}}]{Olson:2017ud}%
  \BibitemOpen
  \bibfield  {author} {\bibinfo {author} {\bibfnamefont {J.}~\bibnamefont
  {Olson}}, \bibinfo {author} {\bibfnamefont {Y.}~\bibnamefont {Cao}}, \bibinfo
  {author} {\bibfnamefont {J.}~\bibnamefont {Romero}}, \bibinfo {author}
  {\bibfnamefont {P.}~\bibnamefont {Johnson}}, \bibinfo {author} {\bibfnamefont
  {P.-L.}\ \bibnamefont {Dallaire-Demers}}, \bibinfo {author} {\bibfnamefont
  {N.}~\bibnamefont {Sawaya}}, \bibinfo {author} {\bibfnamefont
  {P.}~\bibnamefont {Narang}}, \bibinfo {author} {\bibfnamefont
  {I.}~\bibnamefont {Kivlichan}}, \bibinfo {author} {\bibfnamefont
  {M.}~\bibnamefont {Wasielewski}}, \ and\ \bibinfo {author} {\bibfnamefont
  {A.}~\bibnamefont {Aspuru-Guzik}},\ }\href@noop {} {\bibfield  {journal}
  {\bibinfo  {journal} {arXiv.org}\ ,\ \bibinfo {pages} {arXiv:1706.05413v2}}
  (\bibinfo {year} {2017})}\BibitemShut {NoStop}%
\bibitem [{\citenamefont {McArdle}\ \emph {et~al.}(2018)\citenamefont
  {McArdle}, \citenamefont {Endo}, \citenamefont {Aspuru-Guzik}, \citenamefont
  {Benjamin},\ and\ \citenamefont {Yuan}}]{McArdle:2018we}%
  \BibitemOpen
  \bibfield  {author} {\bibinfo {author} {\bibfnamefont {S.}~\bibnamefont
  {McArdle}}, \bibinfo {author} {\bibfnamefont {S.}~\bibnamefont {Endo}},
  \bibinfo {author} {\bibfnamefont {A.}~\bibnamefont {Aspuru-Guzik}}, \bibinfo
  {author} {\bibfnamefont {S.}~\bibnamefont {Benjamin}}, \ and\ \bibinfo
  {author} {\bibfnamefont {X.}~\bibnamefont {Yuan}},\ }\href@noop {} {\bibfield
   {journal} {\bibinfo  {journal} {arXiv.org}\ ,\ \bibinfo {pages}
  {arXiv:1808.10402v1}} (\bibinfo {year} {2018})}\BibitemShut {NoStop}%
\bibitem [{\citenamefont {Helgaker}\ \emph {et~al.}(2000)\citenamefont
  {Helgaker}, \citenamefont {Jorgensen},\ and\ \citenamefont
  {Olsen}}]{Helgaker:2000}%
  \BibitemOpen
  \bibfield  {author} {\bibinfo {author} {\bibfnamefont {T.}~\bibnamefont
  {Helgaker}}, \bibinfo {author} {\bibfnamefont {P.}~\bibnamefont {Jorgensen}},
  \ and\ \bibinfo {author} {\bibfnamefont {J.}~\bibnamefont {Olsen}},\
  }\href@noop {} {\emph {\bibinfo {title} {Molecular Electronic-structure
  Theory}}}\ (\bibinfo  {publisher} {Wiley},\ \bibinfo {year}
  {2000})\BibitemShut {NoStop}%
\bibitem [{\citenamefont {Bravyi}\ and\ \citenamefont
  {Kitaev}(2002)}]{Bravyi:2002/aph/210}%
  \BibitemOpen
  \bibfield  {author} {\bibinfo {author} {\bibfnamefont {S.~B.}\ \bibnamefont
  {Bravyi}}\ and\ \bibinfo {author} {\bibfnamefont {A.~Y.}\ \bibnamefont
  {Kitaev}},\ }\href {\doibase 10.1006/aphy.2002.6254} {\bibfield  {journal}
  {\bibinfo  {journal} {Ann. Phys.}\ }\textbf {\bibinfo {volume} {298}},\
  \bibinfo {pages} {210} (\bibinfo {year} {2002})}\BibitemShut {NoStop}%
\bibitem [{\citenamefont {Seeley}\ \emph {et~al.}(2012)\citenamefont {Seeley},
  \citenamefont {Richard},\ and\ \citenamefont
  {Love}}]{Seeley:2012/jcp/224109}%
  \BibitemOpen
  \bibfield  {author} {\bibinfo {author} {\bibfnamefont {J.~T.}\ \bibnamefont
  {Seeley}}, \bibinfo {author} {\bibfnamefont {M.~J.}\ \bibnamefont {Richard}},
  \ and\ \bibinfo {author} {\bibfnamefont {P.~J.}\ \bibnamefont {Love}},\
  }\href {\doibase 10.1063/1.4768229} {\bibfield  {journal} {\bibinfo
  {journal} {J. Chem. Phys.}\ }\textbf {\bibinfo {volume} {137}},\ \bibinfo
  {pages} {224109} (\bibinfo {year} {2012})}\BibitemShut {NoStop}%
\bibitem [{\citenamefont {Tranter}\ \emph {et~al.}(2015)\citenamefont
  {Tranter}, \citenamefont {Sofia}, \citenamefont {Seeley}, \citenamefont
  {Kaicher}, \citenamefont {McClean}, \citenamefont {Babbush}, \citenamefont
  {Coveney}, \citenamefont {Mintert}, \citenamefont {Wilhelm},\ and\
  \citenamefont {Love}}]{Tranter:2015/ijqc/1431}%
  \BibitemOpen
  \bibfield  {author} {\bibinfo {author} {\bibfnamefont {A.}~\bibnamefont
  {Tranter}}, \bibinfo {author} {\bibfnamefont {S.}~\bibnamefont {Sofia}},
  \bibinfo {author} {\bibfnamefont {J.}~\bibnamefont {Seeley}}, \bibinfo
  {author} {\bibfnamefont {M.}~\bibnamefont {Kaicher}}, \bibinfo {author}
  {\bibfnamefont {J.}~\bibnamefont {McClean}}, \bibinfo {author} {\bibfnamefont
  {R.}~\bibnamefont {Babbush}}, \bibinfo {author} {\bibfnamefont {P.~V.}\
  \bibnamefont {Coveney}}, \bibinfo {author} {\bibfnamefont {F.}~\bibnamefont
  {Mintert}}, \bibinfo {author} {\bibfnamefont {F.}~\bibnamefont {Wilhelm}}, \
  and\ \bibinfo {author} {\bibfnamefont {P.~J.}\ \bibnamefont {Love}},\ }\href
  {\doibase 10.1002/qua.24969} {\bibfield  {journal} {\bibinfo  {journal} {Int.
  J. Quantum Chem.}\ }\textbf {\bibinfo {volume} {115}},\ \bibinfo {pages}
  {1431} (\bibinfo {year} {2015})}\BibitemShut {NoStop}%
\bibitem [{\citenamefont {Setia}\ and\ \citenamefont
  {Whitfield}(2018)}]{Setia:2017/ArXiv/1712.00446}%
  \BibitemOpen
  \bibfield  {author} {\bibinfo {author} {\bibfnamefont {K.}~\bibnamefont
  {Setia}}\ and\ \bibinfo {author} {\bibfnamefont {J.~D.}\ \bibnamefont
  {Whitfield}},\ }\href {\doibase 10.1063/1.5019371} {\bibfield  {journal}
  {\bibinfo  {journal} {J. Chem. Phys.}\ }\textbf {\bibinfo {volume} {148}},\
  \bibinfo {pages} {164104} (\bibinfo {year} {2018})}\BibitemShut {NoStop}%
\bibitem [{\citenamefont {Havl{\'{i}}{\v{c}}ek}\ \emph
  {et~al.}(2017)\citenamefont {Havl{\'{i}}{\v{c}}ek}, \citenamefont {Troyer},\
  and\ \citenamefont {Whitfield}}]{Havlicek:2017/pra/032332}%
  \BibitemOpen
  \bibfield  {author} {\bibinfo {author} {\bibfnamefont {V.}~\bibnamefont
  {Havl{\'{i}}{\v{c}}ek}}, \bibinfo {author} {\bibfnamefont {M.}~\bibnamefont
  {Troyer}}, \ and\ \bibinfo {author} {\bibfnamefont {J.~D.}\ \bibnamefont
  {Whitfield}},\ }\href {\doibase 10.1103/PhysRevA.95.032332} {\bibfield
  {journal} {\bibinfo  {journal} {Phys. Rev. A}\ }\textbf {\bibinfo {volume}
  {95}},\ \bibinfo {pages} {032332} (\bibinfo {year} {2017})}\BibitemShut
  {NoStop}%
\bibitem [{\citenamefont {Cirac}\ and\ \citenamefont
  {Zoller}(2012)}]{cirac:2012}%
  \BibitemOpen
  \bibfield  {author} {\bibinfo {author} {\bibfnamefont {J.~I.}\ \bibnamefont
  {Cirac}}\ and\ \bibinfo {author} {\bibfnamefont {P.}~\bibnamefont {Zoller}},\
  }\href {\doibase 10.1038/nphys2275} {\bibfield  {journal} {\bibinfo
  {journal} {Nat. Phys.}\ }\textbf {\bibinfo {volume} {8}},\ \bibinfo {pages}
  {264} (\bibinfo {year} {2012})}\BibitemShut {NoStop}%
\bibitem [{\citenamefont {Argüello-Luengo}\ \emph {et~al.}(2018)\citenamefont
  {Argüello-Luengo}, \citenamefont {González-Tudela}, \citenamefont {Shi},
  \citenamefont {Zoller},\ and\ \citenamefont {Cirac}}]{cirac:2018}%
  \BibitemOpen
  \bibfield  {author} {\bibinfo {author} {\bibfnamefont {J.}~\bibnamefont
  {Argüello-Luengo}}, \bibinfo {author} {\bibfnamefont {A.}~\bibnamefont
  {González-Tudela}}, \bibinfo {author} {\bibfnamefont {T.}~\bibnamefont
  {Shi}}, \bibinfo {author} {\bibfnamefont {P.}~\bibnamefont {Zoller}}, \ and\
  \bibinfo {author} {\bibfnamefont {J.~I.}\ \bibnamefont {Cirac}},\ }\href@noop
  {} {\bibfield  {journal} {\bibinfo  {journal} {arXiv.org}\ ,\ \bibinfo
  {pages} {arXiv:1807.09228}} (\bibinfo {year} {2018})}\BibitemShut {NoStop}%
\bibitem [{\citenamefont {Verteletskyi}\ \emph {et~al.}(2019)\citenamefont
  {Verteletskyi}, \citenamefont {Yen},\ and\ \citenamefont
  {Izmaylov}}]{VVpap1}%
  \BibitemOpen
  \bibfield  {author} {\bibinfo {author} {\bibfnamefont {V.}~\bibnamefont
  {Verteletskyi}}, \bibinfo {author} {\bibfnamefont {T.-C.}\ \bibnamefont
  {Yen}}, \ and\ \bibinfo {author} {\bibfnamefont {A.~F.}\ \bibnamefont
  {Izmaylov}},\ }\href@noop {} {\bibfield  {journal} {\bibinfo  {journal}
  {arXiv.org}\ ,\ \bibinfo {pages} {arXiv:1907.03358}} (\bibinfo {year}
  {2019})}\BibitemShut {NoStop}%
\bibitem [{\citenamefont {Babbush}\ \emph {et~al.}(2018)\citenamefont
  {Babbush}, \citenamefont {Wiebe}, \citenamefont {McClean}, \citenamefont
  {McClain}, \citenamefont {Neven},\ and\ \citenamefont
  {Chan}}]{Babbush:PRX2018}%
  \BibitemOpen
  \bibfield  {author} {\bibinfo {author} {\bibfnamefont {R.}~\bibnamefont
  {Babbush}}, \bibinfo {author} {\bibfnamefont {N.}~\bibnamefont {Wiebe}},
  \bibinfo {author} {\bibfnamefont {J.}~\bibnamefont {McClean}}, \bibinfo
  {author} {\bibfnamefont {J.}~\bibnamefont {McClain}}, \bibinfo {author}
  {\bibfnamefont {H.}~\bibnamefont {Neven}}, \ and\ \bibinfo {author}
  {\bibfnamefont {G.~K.-L.}\ \bibnamefont {Chan}},\ }\href {\doibase
  10.1103/PhysRevX.8.011044} {\bibfield  {journal} {\bibinfo  {journal} {Phys.
  Rev. X}\ }\textbf {\bibinfo {volume} {8}},\ \bibinfo {pages} {011044}
  (\bibinfo {year} {2018})}\BibitemShut {NoStop}%
\bibitem [{\citenamefont {Columbic}(2004)}]{GraphBook:2004}%
  \BibitemOpen
  \bibfield  {author} {\bibinfo {author} {\bibfnamefont {M.~C.}\ \bibnamefont
  {Columbic}},\ }\href@noop {} {\emph {\bibinfo {title} {{Algorithmic Graph
  Theory and Perfect Graphs}}}},\ Vol.~\bibinfo {volume} {54}\ (\bibinfo
  {publisher} {Annals of Discrete Mathematics (Elsevier)},\ \bibinfo {year}
  {2004})\BibitemShut {NoStop}%
\bibitem [{\citenamefont {Karp}(1972)}]{Karp:1972}%
  \BibitemOpen
  \bibfield  {author} {\bibinfo {author} {\bibfnamefont {R.~M.}\ \bibnamefont
  {Karp}},\ }in\ \href {\doibase 10.1007/978-1-4684-2001-2_9} {\emph {\bibinfo
  {booktitle} {Complexity of Computer Computations. The IBM Research Symposia
  Series}}},\ \bibinfo {editor} {edited by\ \bibinfo {editor} {\bibfnamefont
  {R.}~\bibnamefont {Miller}}, \bibinfo {editor} {\bibfnamefont
  {J.}~\bibnamefont {Thatcher}}, \ and\ \bibinfo {editor} {\bibfnamefont
  {J.}~\bibnamefont {Bohlinger}}}\ (\bibinfo  {publisher} {Springer, Boston,
  MA},\ \bibinfo {year} {1972})\ p.\ \bibinfo {pages} {85–103}\BibitemShut
  {NoStop}%
\bibitem [{\citenamefont {Izmaylov}\ \emph {et~al.}(2019)\citenamefont
  {Izmaylov}, \citenamefont {Yen},\ and\ \citenamefont
  {Ryabinkin}}]{Izmaylov:2019gb}%
  \BibitemOpen
  \bibfield  {author} {\bibinfo {author} {\bibfnamefont {A.~F.}\ \bibnamefont
  {Izmaylov}}, \bibinfo {author} {\bibfnamefont {T.-C.}\ \bibnamefont {Yen}}, \
  and\ \bibinfo {author} {\bibfnamefont {I.~G.}\ \bibnamefont {Ryabinkin}},\
  }\href@noop {} {\bibfield  {journal} {\bibinfo  {journal} {Chem. Sci.}\
  }\textbf {\bibinfo {volume} {10}},\ \bibinfo {pages} {3746} (\bibinfo {year}
  {2019})}\BibitemShut {NoStop}%
\bibitem [{\citenamefont {Albarr{\'a}n-Arriagada}\ \emph
  {et~al.}(2018)\citenamefont {Albarr{\'a}n-Arriagada}, \citenamefont
  {Barrios}, \citenamefont {Sanz}, \citenamefont {Romero}, \citenamefont
  {Lamata}, \citenamefont {Retamal},\ and\ \citenamefont
  {Solano}}]{nArriagada:2018ju}%
  \BibitemOpen
  \bibfield  {author} {\bibinfo {author} {\bibfnamefont {F.}~\bibnamefont
  {Albarr{\'a}n-Arriagada}}, \bibinfo {author} {\bibfnamefont {G.~A.}\
  \bibnamefont {Barrios}}, \bibinfo {author} {\bibfnamefont {M.}~\bibnamefont
  {Sanz}}, \bibinfo {author} {\bibfnamefont {G.}~\bibnamefont {Romero}},
  \bibinfo {author} {\bibfnamefont {L.}~\bibnamefont {Lamata}}, \bibinfo
  {author} {\bibfnamefont {J.~C.}\ \bibnamefont {Retamal}}, \ and\ \bibinfo
  {author} {\bibfnamefont {E.}~\bibnamefont {Solano}},\ }\href@noop {}
  {\bibfield  {journal} {\bibinfo  {journal} {Phys. Rev. A}\ }\textbf {\bibinfo
  {volume} {97}},\ \bibinfo {pages} {032320:1} (\bibinfo {year}
  {2018})}\BibitemShut {NoStop}%
\bibitem [{\citenamefont {Prevedel}\ \emph {et~al.}(2007)\citenamefont
  {Prevedel}, \citenamefont {Walther}, \citenamefont {Tiefenbacher},
  \citenamefont {B{\"o}hi}, \citenamefont {Kaltenbaek}, \citenamefont
  {Jennewein},\ and\ \citenamefont {Zeilinger}}]{Prevedel:2007ca}%
  \BibitemOpen
  \bibfield  {author} {\bibinfo {author} {\bibfnamefont {R.}~\bibnamefont
  {Prevedel}}, \bibinfo {author} {\bibfnamefont {P.}~\bibnamefont {Walther}},
  \bibinfo {author} {\bibfnamefont {F.}~\bibnamefont {Tiefenbacher}}, \bibinfo
  {author} {\bibfnamefont {P.}~\bibnamefont {B{\"o}hi}}, \bibinfo {author}
  {\bibfnamefont {R.}~\bibnamefont {Kaltenbaek}}, \bibinfo {author}
  {\bibfnamefont {T.}~\bibnamefont {Jennewein}}, \ and\ \bibinfo {author}
  {\bibfnamefont {A.}~\bibnamefont {Zeilinger}},\ }\href@noop {} {\bibfield
  {journal} {\bibinfo  {journal} {Nature}\ }\textbf {\bibinfo {volume} {445}},\
  \bibinfo {pages} {65} (\bibinfo {year} {2007})}\BibitemShut {NoStop}%
\bibitem [{\citenamefont {Procopio}\ \emph {et~al.}(2015)\citenamefont
  {Procopio}, \citenamefont {Moqanaki}, \citenamefont {Ara{\'u}jo},
  \citenamefont {Costa}, \citenamefont {Calafell}, \citenamefont {Dowd},
  \citenamefont {Hamel}, \citenamefont {Rozema}, \citenamefont {Brukner},\ and\
  \citenamefont {Walther}}]{Moqanaki:2015iw}%
  \BibitemOpen
  \bibfield  {author} {\bibinfo {author} {\bibfnamefont {L.~M.}\ \bibnamefont
  {Procopio}}, \bibinfo {author} {\bibfnamefont {A.}~\bibnamefont {Moqanaki}},
  \bibinfo {author} {\bibfnamefont {M.}~\bibnamefont {Ara{\'u}jo}}, \bibinfo
  {author} {\bibfnamefont {F.}~\bibnamefont {Costa}}, \bibinfo {author}
  {\bibfnamefont {I.~A.}\ \bibnamefont {Calafell}}, \bibinfo {author}
  {\bibfnamefont {E.~G.}\ \bibnamefont {Dowd}}, \bibinfo {author}
  {\bibfnamefont {D.~R.}\ \bibnamefont {Hamel}}, \bibinfo {author}
  {\bibfnamefont {L.~A.}\ \bibnamefont {Rozema}}, \bibinfo {author}
  {\bibfnamefont {v.}~\bibnamefont {Brukner}}, \ and\ \bibinfo {author}
  {\bibfnamefont {P.}~\bibnamefont {Walther}},\ }\href@noop {} {\bibfield
  {journal} {\bibinfo  {journal} {Nat. Commun.}\ }\textbf {\bibinfo {volume}
  {6}},\ \bibinfo {pages} {7913:1} (\bibinfo {year} {2015})}\BibitemShut
  {NoStop}%
\bibitem [{\citenamefont {Reimer}\ \emph {et~al.}(2019)\citenamefont {Reimer},
  \citenamefont {Sciara}, \citenamefont {Roztocki}, \citenamefont {Islam},
  \citenamefont {Cort{\'e}s}, \citenamefont {Zhang}, \citenamefont {Fischer},
  \citenamefont {Loranger}, \citenamefont {Kashyap}, \citenamefont {Cino},
  \citenamefont {Chu}, \citenamefont {Little}, \citenamefont {Moss},
  \citenamefont {Caspani}, \citenamefont {Munro}, \citenamefont {Aza{\~n}a},
  \citenamefont {Kues},\ and\ \citenamefont {Morandotti}}]{Reimer:2018cv}%
  \BibitemOpen
  \bibfield  {author} {\bibinfo {author} {\bibfnamefont {C.}~\bibnamefont
  {Reimer}}, \bibinfo {author} {\bibfnamefont {S.}~\bibnamefont {Sciara}},
  \bibinfo {author} {\bibfnamefont {P.}~\bibnamefont {Roztocki}}, \bibinfo
  {author} {\bibfnamefont {M.}~\bibnamefont {Islam}}, \bibinfo {author}
  {\bibfnamefont {L.~R.}\ \bibnamefont {Cort{\'e}s}}, \bibinfo {author}
  {\bibfnamefont {Y.}~\bibnamefont {Zhang}}, \bibinfo {author} {\bibfnamefont
  {B.}~\bibnamefont {Fischer}}, \bibinfo {author} {\bibfnamefont
  {S.}~\bibnamefont {Loranger}}, \bibinfo {author} {\bibfnamefont
  {R.}~\bibnamefont {Kashyap}}, \bibinfo {author} {\bibfnamefont
  {A.}~\bibnamefont {Cino}}, \bibinfo {author} {\bibfnamefont {S.~T.}\
  \bibnamefont {Chu}}, \bibinfo {author} {\bibfnamefont {B.~E.}\ \bibnamefont
  {Little}}, \bibinfo {author} {\bibfnamefont {D.~J.}\ \bibnamefont {Moss}},
  \bibinfo {author} {\bibfnamefont {L.}~\bibnamefont {Caspani}}, \bibinfo
  {author} {\bibfnamefont {W.~J.}\ \bibnamefont {Munro}}, \bibinfo {author}
  {\bibfnamefont {J.}~\bibnamefont {Aza{\~n}a}}, \bibinfo {author}
  {\bibfnamefont {M.}~\bibnamefont {Kues}}, \ and\ \bibinfo {author}
  {\bibfnamefont {R.}~\bibnamefont {Morandotti}},\ }\href@noop {} {\bibfield
  {journal} {\bibinfo  {journal} {Nat. Phys.}\ }\textbf {\bibinfo {volume}
  {15}},\ \bibinfo {pages} {148} (\bibinfo {year} {2019})}\BibitemShut
  {NoStop}%
\bibitem [{\citenamefont {Buhrman}\ \emph {et~al.}(2001)\citenamefont
  {Buhrman}, \citenamefont {Cleve}, \citenamefont {Watrous},\ and\
  \citenamefont {de~Wolf}}]{PhysRevLett.87.167902}%
  \BibitemOpen
  \bibfield  {author} {\bibinfo {author} {\bibfnamefont {H.}~\bibnamefont
  {Buhrman}}, \bibinfo {author} {\bibfnamefont {R.}~\bibnamefont {Cleve}},
  \bibinfo {author} {\bibfnamefont {J.}~\bibnamefont {Watrous}}, \ and\
  \bibinfo {author} {\bibfnamefont {R.}~\bibnamefont {de~Wolf}},\ }\href
  {\doibase 10.1103/PhysRevLett.87.167902} {\bibfield  {journal} {\bibinfo
  {journal} {Phys. Rev. Lett.}\ }\textbf {\bibinfo {volume} {87}},\ \bibinfo
  {pages} {167902} (\bibinfo {year} {2001})}\BibitemShut {NoStop}%
\bibitem [{\citenamefont {Ryabinkin}\ \emph {et~al.}(2018)\citenamefont
  {Ryabinkin}, \citenamefont {Yen}, \citenamefont {Genin},\ and\ \citenamefont
  {Izmaylov}}]{Ryabinkin:2018/qcc}%
  \BibitemOpen
  \bibfield  {author} {\bibinfo {author} {\bibfnamefont {I.~G.}\ \bibnamefont
  {Ryabinkin}}, \bibinfo {author} {\bibfnamefont {T.-C.}\ \bibnamefont {Yen}},
  \bibinfo {author} {\bibfnamefont {S.~N.}\ \bibnamefont {Genin}}, \ and\
  \bibinfo {author} {\bibfnamefont {A.~F.}\ \bibnamefont {Izmaylov}},\
  }\href@noop {} {\bibfield  {journal} {\bibinfo  {journal} {J. Chem. Theory
  Comput.}\ }\textbf {\bibinfo {volume} {14}},\ \bibinfo {pages} {6317}
  (\bibinfo {year} {2018})}\BibitemShut {NoStop}%
\bibitem [{\citenamefont {Vidal}\ and\ \citenamefont
  {Dawson}(2004)}]{PhysRevA.69.010301}%
  \BibitemOpen
  \bibfield  {author} {\bibinfo {author} {\bibfnamefont {G.}~\bibnamefont
  {Vidal}}\ and\ \bibinfo {author} {\bibfnamefont {C.~M.}\ \bibnamefont
  {Dawson}},\ }\href {\doibase 10.1103/PhysRevA.69.010301} {\bibfield
  {journal} {\bibinfo  {journal} {Phys. Rev. A}\ }\textbf {\bibinfo {volume}
  {69}},\ \bibinfo {pages} {010301} (\bibinfo {year} {2004})}\BibitemShut
  {NoStop}%
\bibitem [{\citenamefont {Yen}\ \emph {et~al.}(2019{\natexlab{a}})\citenamefont
  {Yen}, \citenamefont {Lang},\ and\ \citenamefont {Izmaylov}}]{Yen:2019prj}%
  \BibitemOpen
  \bibfield  {author} {\bibinfo {author} {\bibfnamefont {T.-C.}\ \bibnamefont
  {Yen}}, \bibinfo {author} {\bibfnamefont {R.~A.}\ \bibnamefont {Lang}}, \
  and\ \bibinfo {author} {\bibfnamefont {A.~F.}\ \bibnamefont {Izmaylov}},\
  }\href@noop {} {\bibfield  {journal} {\bibinfo  {journal} {arXiv.org}\ ,\
  \bibinfo {pages} {arXiv:1905.08109v1}} (\bibinfo {year}
  {2019}{\natexlab{a}})}\BibitemShut {NoStop}%
\bibitem [{\citenamefont {Kandala}\ \emph {et~al.}(2017)\citenamefont
  {Kandala}, \citenamefont {Mezzacapo}, \citenamefont {Temme}, \citenamefont
  {Takita}, \citenamefont {Brink}, \citenamefont {Chow},\ and\ \citenamefont
  {Gambetta}}]{Kandala:2017/nature/242}%
  \BibitemOpen
  \bibfield  {author} {\bibinfo {author} {\bibfnamefont {A.}~\bibnamefont
  {Kandala}}, \bibinfo {author} {\bibfnamefont {A.}~\bibnamefont {Mezzacapo}},
  \bibinfo {author} {\bibfnamefont {K.}~\bibnamefont {Temme}}, \bibinfo
  {author} {\bibfnamefont {M.}~\bibnamefont {Takita}}, \bibinfo {author}
  {\bibfnamefont {M.}~\bibnamefont {Brink}}, \bibinfo {author} {\bibfnamefont
  {J.~M.}\ \bibnamefont {Chow}}, \ and\ \bibinfo {author} {\bibfnamefont
  {J.~M.}\ \bibnamefont {Gambetta}},\ }\href {\doibase 10.1038/nature23879}
  {\bibfield  {journal} {\bibinfo  {journal} {Nature}\ }\textbf {\bibinfo
  {volume} {549}},\ \bibinfo {pages} {242} (\bibinfo {year}
  {2017})}\BibitemShut {NoStop}%
\bibitem [{\citenamefont {Hempel}\ \emph {et~al.}(2018)\citenamefont {Hempel},
  \citenamefont {Maier}, \citenamefont {Romero}, \citenamefont {McClean},
  \citenamefont {Monz}, \citenamefont {Shen}, \citenamefont {Jurcevic},
  \citenamefont {Lanyon}, \citenamefont {Love}, \citenamefont {Babbush},
  \citenamefont {Aspuru-Guzik}, \citenamefont {Blatt},\ and\ \citenamefont
  {Roos}}]{Hempel:2018/prx/031022}%
  \BibitemOpen
  \bibfield  {author} {\bibinfo {author} {\bibfnamefont {C.}~\bibnamefont
  {Hempel}}, \bibinfo {author} {\bibfnamefont {C.}~\bibnamefont {Maier}},
  \bibinfo {author} {\bibfnamefont {J.}~\bibnamefont {Romero}}, \bibinfo
  {author} {\bibfnamefont {J.}~\bibnamefont {McClean}}, \bibinfo {author}
  {\bibfnamefont {T.}~\bibnamefont {Monz}}, \bibinfo {author} {\bibfnamefont
  {H.}~\bibnamefont {Shen}}, \bibinfo {author} {\bibfnamefont {P.}~\bibnamefont
  {Jurcevic}}, \bibinfo {author} {\bibfnamefont {B.~P.}\ \bibnamefont
  {Lanyon}}, \bibinfo {author} {\bibfnamefont {P.}~\bibnamefont {Love}},
  \bibinfo {author} {\bibfnamefont {R.}~\bibnamefont {Babbush}}, \bibinfo
  {author} {\bibfnamefont {A.}~\bibnamefont {Aspuru-Guzik}}, \bibinfo {author}
  {\bibfnamefont {R.}~\bibnamefont {Blatt}}, \ and\ \bibinfo {author}
  {\bibfnamefont {C.~F.}\ \bibnamefont {Roos}},\ }\href {\doibase
  10.1103/PhysRevX.8.031022} {\bibfield  {journal} {\bibinfo  {journal} {Phys.
  Rev. X}\ }\textbf {\bibinfo {volume} {8}},\ \bibinfo {pages} {031022}
  (\bibinfo {year} {2018})}\BibitemShut {NoStop}%
\bibitem [{\citenamefont {{Rigetti\ Computing}}(2018)}]{Rigetti_doc}%
  \BibitemOpen
  \bibfield  {author} {\bibinfo {author} {\bibnamefont {{Rigetti\
  Computing}}},\ }\href@noop {} {\enquote {\bibinfo {title} {{pyQuil 1.9}},}\ }
  (\bibinfo {year} {2018}),\ \bibinfo {note}
  {http://docs.rigetti.com/en/1.9/qpu.html}\BibitemShut {NoStop}%
\bibitem [{\citenamefont {Welsh}(1967)}]{Welsh:1967}%
  \BibitemOpen
  \bibfield  {author} {\bibinfo {author} {\bibfnamefont {D.~J.~A.}\
  \bibnamefont {Welsh}},\ }\href {\doibase 10.1093/comjnl/10.1.85} {\bibfield
  {journal} {\bibinfo  {journal} {Comput. J.}\ }\textbf {\bibinfo {volume}
  {10}},\ \bibinfo {pages} {85–86} (\bibinfo {year} {1967})}\BibitemShut
  {NoStop}%
\bibitem [{\citenamefont {Matula}\ \emph {et~al.}(1972)\citenamefont {Matula},
  \citenamefont {Marble},\ and\ \citenamefont {Isaacson}}]{Matula:1972}%
  \BibitemOpen
  \bibfield  {author} {\bibinfo {author} {\bibfnamefont {D.~W.}\ \bibnamefont
  {Matula}}, \bibinfo {author} {\bibfnamefont {G.}~\bibnamefont {Marble}}, \
  and\ \bibinfo {author} {\bibfnamefont {J.~D.}\ \bibnamefont {Isaacson}},\
  }in\ \href {\doibase 10.1016/b978-1-4832-3187-7.50015-5} {\emph {\bibinfo
  {booktitle} {Graph Theory and Computing}}},\ \bibinfo {editor} {edited by\
  \bibinfo {editor} {\bibfnamefont {R.~C.}\ \bibnamefont {Read}}}\ (\bibinfo
  {publisher} {Academic Press},\ \bibinfo {year} {1972})\ pp.\ \bibinfo {pages}
  {109 -- 122}\BibitemShut {NoStop}%
\bibitem [{\citenamefont {Brélaz}(1979)}]{Brelaz:1973}%
  \BibitemOpen
  \bibfield  {author} {\bibinfo {author} {\bibfnamefont {D.}~\bibnamefont
  {Brélaz}},\ }\href {\doibase 10.1145/359094.359101} {\bibfield  {journal}
  {\bibinfo  {journal} {Commun. ACM}\ }\textbf {\bibinfo {volume} {22}},\
  \bibinfo {pages} {251–256} (\bibinfo {year} {1979})}\BibitemShut {NoStop}%
\bibitem [{\citenamefont {Leighton}(1979)}]{Leighton:1979}%
  \BibitemOpen
  \bibfield  {author} {\bibinfo {author} {\bibfnamefont {F.~T.}\ \bibnamefont
  {Leighton}},\ }\href@noop {} {\bibfield  {journal} {\bibinfo  {journal} {J.
  Res. Natl. Bur. Stand.}\ }\textbf {\bibinfo {volume} {84}},\ \bibinfo {pages}
  {489} (\bibinfo {year} {1979})}\BibitemShut {NoStop}%
\bibitem [{\citenamefont {Dutton}\ and\ \citenamefont
  {Brigham}(1981)}]{dutton_brigham_1981}%
  \BibitemOpen
  \bibfield  {author} {\bibinfo {author} {\bibfnamefont {R.~D.}\ \bibnamefont
  {Dutton}}\ and\ \bibinfo {author} {\bibfnamefont {R.~C.}\ \bibnamefont
  {Brigham}},\ }\href {\doibase 10.1093/comjnl/24.1.85} {\bibfield  {journal}
  {\bibinfo  {journal} {Comput. J.}\ }\textbf {\bibinfo {volume} {24}},\
  \bibinfo {pages} {85–86} (\bibinfo {year} {1981})}\BibitemShut {NoStop}%
\bibitem [{\citenamefont {Hertz}(1990)}]{hertz_1990}%
  \BibitemOpen
  \bibfield  {author} {\bibinfo {author} {\bibfnamefont {A.}~\bibnamefont
  {Hertz}},\ }\href {\doibase 10.1016/0095-8956(90)90078-e} {\bibfield
  {journal} {\bibinfo  {journal} {J. Comb. Theory}\ }\textbf {\bibinfo {volume}
  {50}},\ \bibinfo {pages} {231–240} (\bibinfo {year} {1990})}\BibitemShut
  {NoStop}%
\bibitem [{\citenamefont {Boppana}\ and\ \citenamefont
  {Halldórsson}(1992)}]{boppana}%
  \BibitemOpen
  \bibfield  {author} {\bibinfo {author} {\bibfnamefont {R.}~\bibnamefont
  {Boppana}}\ and\ \bibinfo {author} {\bibfnamefont {M.~M.}\ \bibnamefont
  {Halldórsson}},\ }\href {\doibase 10.1007/bf01994876} {\bibfield  {journal}
  {\bibinfo  {journal} {BIT Numer. Math}\ }\textbf {\bibinfo {volume} {32}},\
  \bibinfo {pages} {180–196} (\bibinfo {year} {1992})}\BibitemShut {NoStop}%
\bibitem [{\citenamefont {Tomita}\ \emph {et~al.}(2006)\citenamefont {Tomita},
  \citenamefont {Tanaka},\ and\ \citenamefont
  {Takahashi}}]{tomita_tanaka_takahashi_2006}%
  \BibitemOpen
  \bibfield  {author} {\bibinfo {author} {\bibfnamefont {E.}~\bibnamefont
  {Tomita}}, \bibinfo {author} {\bibfnamefont {A.}~\bibnamefont {Tanaka}}, \
  and\ \bibinfo {author} {\bibfnamefont {H.}~\bibnamefont {Takahashi}},\ }\href
  {\doibase 10.1016/j.tcs.2006.06.015} {\bibfield  {journal} {\bibinfo
  {journal} {Theor. Comput. Sci.}\ }\textbf {\bibinfo {volume} {363}},\
  \bibinfo {pages} {28–42} (\bibinfo {year} {2006})}\BibitemShut {NoStop}%
\bibitem [{\citenamefont {Zhao}\ \emph {et~al.}(2019)\citenamefont {Zhao},
  \citenamefont {Tranter}, \citenamefont {Kirby}, \citenamefont {Ung},
  \citenamefont {Miyake},\ and\ \citenamefont {Love}}]{Zhao:2019vz}%
  \BibitemOpen
  \bibfield  {author} {\bibinfo {author} {\bibfnamefont {A.}~\bibnamefont
  {Zhao}}, \bibinfo {author} {\bibfnamefont {A.}~\bibnamefont {Tranter}},
  \bibinfo {author} {\bibfnamefont {W.~M.}\ \bibnamefont {Kirby}}, \bibinfo
  {author} {\bibfnamefont {S.~F.}\ \bibnamefont {Ung}}, \bibinfo {author}
  {\bibfnamefont {A.}~\bibnamefont {Miyake}}, \ and\ \bibinfo {author}
  {\bibfnamefont {P.}~\bibnamefont {Love}},\ }\href@noop {} {\bibfield
  {journal} {\bibinfo  {journal} {arXiv.org}\ ,\ \bibinfo {pages}
  {arXiv:1908.08067v1}} (\bibinfo {year} {2019})}\BibitemShut {NoStop}%
\bibitem [{\citenamefont {Jena}\ \emph {et~al.}(2019)\citenamefont {Jena},
  \citenamefont {Genin},\ and\ \citenamefont {Mosca}}]{MoscaA}%
  \BibitemOpen
  \bibfield  {author} {\bibinfo {author} {\bibfnamefont {A.}~\bibnamefont
  {Jena}}, \bibinfo {author} {\bibfnamefont {S.}~\bibnamefont {Genin}}, \ and\
  \bibinfo {author} {\bibfnamefont {M.}~\bibnamefont {Mosca}},\ }\href@noop {}
  {\bibfield  {journal} {\bibinfo  {journal} {arXiv.org}\ ,\ \bibinfo {pages}
  {arXiv:1907.07859}} (\bibinfo {year} {2019})}\BibitemShut {NoStop}%
\bibitem [{\citenamefont {Yen}\ \emph {et~al.}(2019{\natexlab{b}})\citenamefont
  {Yen}, \citenamefont {Verteletskyi},\ and\ \citenamefont
  {Izmaylov}}]{ThomsonA}%
  \BibitemOpen
  \bibfield  {author} {\bibinfo {author} {\bibfnamefont {T.-C.}\ \bibnamefont
  {Yen}}, \bibinfo {author} {\bibfnamefont {V.}~\bibnamefont {Verteletskyi}}, \
  and\ \bibinfo {author} {\bibfnamefont {A.~F.}\ \bibnamefont {Izmaylov}},\
  }\href@noop {} {\bibfield  {journal} {\bibinfo  {journal} {arXiv.org}\ ,\
  \bibinfo {pages} {arXiv:1907.09386}} (\bibinfo {year}
  {2019}{\natexlab{b}})}\BibitemShut {NoStop}%
\bibitem [{\citenamefont {Huggins}\ \emph {et~al.}(2019)\citenamefont
  {Huggins}, \citenamefont {McClean}, \citenamefont {Rubin}, \citenamefont
  {Jiang}, \citenamefont {Wiebe}, \citenamefont {Whaley},\ and\ \citenamefont
  {Babbush}}]{BabbushA}%
  \BibitemOpen
  \bibfield  {author} {\bibinfo {author} {\bibfnamefont {W.~J.}\ \bibnamefont
  {Huggins}}, \bibinfo {author} {\bibfnamefont {J.}~\bibnamefont {McClean}},
  \bibinfo {author} {\bibfnamefont {N.}~\bibnamefont {Rubin}}, \bibinfo
  {author} {\bibfnamefont {Z.}~\bibnamefont {Jiang}}, \bibinfo {author}
  {\bibfnamefont {N.}~\bibnamefont {Wiebe}}, \bibinfo {author} {\bibfnamefont
  {K.~B.}\ \bibnamefont {Whaley}}, \ and\ \bibinfo {author} {\bibfnamefont
  {R.}~\bibnamefont {Babbush}},\ }\href@noop {} {\bibfield  {journal} {\bibinfo
   {journal} {arXiv.org}\ ,\ \bibinfo {pages} {arXiv:1907.13117}} (\bibinfo
  {year} {2019})}\BibitemShut {NoStop}%
\bibitem [{\citenamefont {Gokhale}\ \emph {et~al.}(2019)\citenamefont
  {Gokhale}, \citenamefont {Angiuli}, \citenamefont {Ding}, \citenamefont
  {Gui}, \citenamefont {Tomesh}, \citenamefont {Suchara}, \citenamefont
  {Martonosi},\ and\ \citenamefont {Chong}}]{ChicagoA}%
  \BibitemOpen
  \bibfield  {author} {\bibinfo {author} {\bibfnamefont {P.}~\bibnamefont
  {Gokhale}}, \bibinfo {author} {\bibfnamefont {O.}~\bibnamefont {Angiuli}},
  \bibinfo {author} {\bibfnamefont {Y.}~\bibnamefont {Ding}}, \bibinfo {author}
  {\bibfnamefont {K.}~\bibnamefont {Gui}}, \bibinfo {author} {\bibfnamefont
  {T.}~\bibnamefont {Tomesh}}, \bibinfo {author} {\bibfnamefont
  {M.}~\bibnamefont {Suchara}}, \bibinfo {author} {\bibfnamefont
  {M.}~\bibnamefont {Martonosi}}, \ and\ \bibinfo {author} {\bibfnamefont
  {F.~T.}\ \bibnamefont {Chong}},\ }\href@noop {} {\bibfield  {journal}
  {\bibinfo  {journal} {arXiv.org}\ ,\ \bibinfo {pages} {arXiv:1907.13623}}
  (\bibinfo {year} {2019})}\BibitemShut {NoStop}%
\end{thebibliography}
%

\end{document}